\documentclass[prl,nopacs,twocolumn,aps]{revtex4}

\pdfoutput=1

\usepackage{color}

\usepackage{amsmath}
\usepackage{bm}
\usepackage{epsfig}
\usepackage{amssymb}

\usepackage{graphicx}
\usepackage{epsfig}
\usepackage{psfrag}

\begin{document}

\title{Thermodynamics of Quantum Jump Trajectories}

\author{Juan P. Garrahan}

\author{Igor Lesanovsky}

\affiliation{School of Physics and Astronomy, University of
Nottingham, Nottingham, NG7 2RD, UK}

\begin{abstract}
We apply the large-deviation method to study trajectories in dissipative quantum systems. We show that in the long time limit the statistics of quantum jumps can be understood from thermodynamic arguments by exploiting the analogy between large-deviation and free-energy functions. This approach is particularly useful for uncovering properties of rare dissipative trajectories. We also prove, via an explicit quantum mapping, that rare trajectories of one system can be realized as typical trajectories of an alternative system.
\end{abstract}


\maketitle

\noindent
{\em Introduction.}
Equilibrium statistical mechanics provides the tools to study equilibrium phases and phase changes in many body systems \cite{Chandler}.  Thermodynamic phases are characterized by average values of thermodynamic observables, such as volume in a liquid or magnetization in a magnet, which are controlled by conjugate fields, such as pressure or magnetic field.  Non-analyticities in free-energies correspond to phase transition points, and the proximity to a phase transition manifests in large and rare fluctuations of observables around their thermodynamic values \cite{Chandler}. 

An analogous perspective can be adopted for the study of {\em dynamical} phases in non-equilibrium systems by applying the large-deviation (LD) method \cite{LD}.  The LD formalism allows to treat ensembles of trajectories, classified by dynamical order parameters or their conjugate fields, in the same way that equilibrium statistical mechanics treats ensembles of configurations.  Important properties of classical non-equilibrium systems can be uncovered by exploiting this analogy \cite{LD,LD-examples,Lecomte}, such as the existence of  ``space-time'' phase transitions in glassy systems \cite{Garrahan}.  

In this Letter we apply the LD method to quantum non-equilibrium systems.  This approach reveals important properties of ensembles of trajectories of quantum systems that undergo quantum jumps in some form, such as driven quantum systems weakly coupled to a thermal bath \cite{Zoller,Plenio}.   We show that one can observe features of dynamical crossovers and dynamical phase transitions even in quantum systems with only a few degrees of freedom, and illustrate our ideas with three simple examples: (i) a driven 2-level system, where the LD approach allows to identify a scale invariance point in the ensemble of trajectories of emitted photons; (ii) a blinking 3-level system (or electron shelving problem), where we argue that intermittency in photon count is related to a crossover between distinct dynamical phases; and (iii) a micromaser, where static bistability leads to a first-order phase transition in the ensemble of trajectories.  We also establish a mapping between two dynamical systems, where typical trajectories of one are the rare trajectories of the other.  This method is particularly useful for generating rare trajectories which otherwise are highly suppressed.

\noindent
{\em Formalism.} We consider a quantum system weakly coupled to a reservoir in the Markovian regime.  The non-unitary evolution of its density matrix $\rho(t)$ is described by a so-called Master equation \cite{Lindblad,Gardiner},
\begin{equation}
\frac{d}{dt}  \rho(t) = -i [ H , \rho] + \sum_{\mu=1}^{N_{\rm L}} \left( L_\mu \rho L_\mu^\dagger - \frac{1}{2} \{  L_\mu^\dagger L_\mu , \rho \} \right) ,
\label{Lindblad}
\end{equation}
where $L_\mu$ and $L_\mu^\dagger$ ($\mu = 1, \ldots, N_{\rm L}$) are the Lindblad operators \cite{Lindblad,Gardiner}, $\{  \cdot , \cdot \}$ stands for anticommutator, and we have set $\hbar = 1$.  We are interested in the time record of projection events due to one (or more) of the Lindblad operators, such as that of emitted photons, which we assume are detected with $100\%$ efficiency.  Such record is a particular quantum jump trajectory of the system \cite{Plenio,Gardiner}.  The probability $P_t(K)$ to observe $K$ events after time $t$ is given by $P_t(K) = {\rm Tr} \left[ \rho^{(K)}(t) \right]$, where $\rho^{(K)}(t)$ is a reduced density matrix  obtained by the projection of the full density matrix onto the subspace of $K$ events, e.g., the subspace containing $K$ photons \cite{Zoller}.   
For large times $P_t(K)$ acquires a LD form:
\begin{equation}
P_t(K) = {\rm Tr}  \left[ \rho^{(K)}(t) \right] \approx e^{-t \varphi(K/t)} .
\label{phi}
\end{equation}
The ``large-deviation'' function $\varphi(k)$ ($k \equiv K/t$) contains all information about the probability of $K$ at long times.  Alternatively, we can describe the statistics of $K$ via the generating function, which also has a LD form,
\begin{equation}
Z_t(s) \equiv \sum_{K=0}^{\infty} P_t(K) e^{-s K} \approx e^{t \theta(s)} .
\label{theta}
\end{equation}
The LD functions $\varphi(k)$ and $\theta(s)$ are to trajectories what entropy density and free-energy density are to configurations in equilibrium statistical mechanics, with $s$ being the conjugate field to the {\em dynamical order parameter} $K$.  The two are related by a Legendre transform, $\theta(s) = - \min_k \left[ \varphi(k) + k s \right]$, and the function $\theta(s)$ has the convexity properties of (minus) a free-energy.  Moreover, anomalous dependence of $\theta(s)$ on $s$ indicates non-trivial fluctuation properties of dynamical trajectories.  In particular, singularities in $\theta(s)$ correspond to dynamical (or space-time \cite{Garrahan}) phase transitions.  It is this anomalous and phase-transition behavior that we uncover below by calculating $\theta(s)$ for simple driven quantum systems. 

The $\rho^{(K)}(t)$ obey a set of equations \cite{Zoller} which is uncoupled by the Laplace transform (\ref{theta}).  That is, the equation for $\rho_s(t) \equiv \sum_{K=0}^{\infty} \rho^{(K)}(t) e^{-s K}$ reads $\frac{d}{dt}  \rho_s(t) = {\cal W}_s(\rho_s)$, where the super-operator ${\cal W}_s$ is 
\begin{eqnarray}
{\cal W}_s(\rho) &=& -i [ H , \rho] + e^{-s} L_1 \rho L_1^\dagger
\nonumber \\
&& + \sum_{\mu=2}^{N_{\rm L}} L_\mu \rho L_\mu^\dagger - \frac{1}{2} \sum_{\mu=1}^{N_{\rm L}} \{  L_\mu^\dagger L_\mu , \rho \} .
\label{Ws}
\end{eqnarray}
$L_1$ is the Lindblad operator which produces the quantum jumps we are counting in $K$.  The equation $\partial_t \rho ={\cal W}_s(\rho)$ is sometimes called the generalized quantum master equation \cite{Brown,Mukamel}. The operator ${\cal W}_s$ is analogous to the Lebowitz-Spohn operator \cite{Lebowitz} of classical non-equilibrium dynamics.  Physical dynamics takes place at $s=0$ [here ${\cal W}_{s} = {\cal W}$, see Eqs.\ (\ref{Lindblad},\ref{Ws})].  For $s \neq 0$, $\partial_t \rho = {\cal W}_s(\rho)$ describes a time evolution whose unfolding \cite{Zoller,Plenio,Belavkin} generates an ensemble of trajectories biased by $e^{-s K}$, see Eq.\ (\ref{theta}).  We call this the $s$-ensemble \cite{Garrahan}.  

In analogy to the classical case we assume that the LD function $\theta(s)$ is given by the largest real eigenvalue of ${\cal W}_s$ \cite{Lebowitz}. This assumption, corroborated below, reduces the problem of calculating the ``partition sum'' of Eq.\ (\ref{theta}) to an eigenvalue problem.

\begin{figure}
\includegraphics[width= \columnwidth]{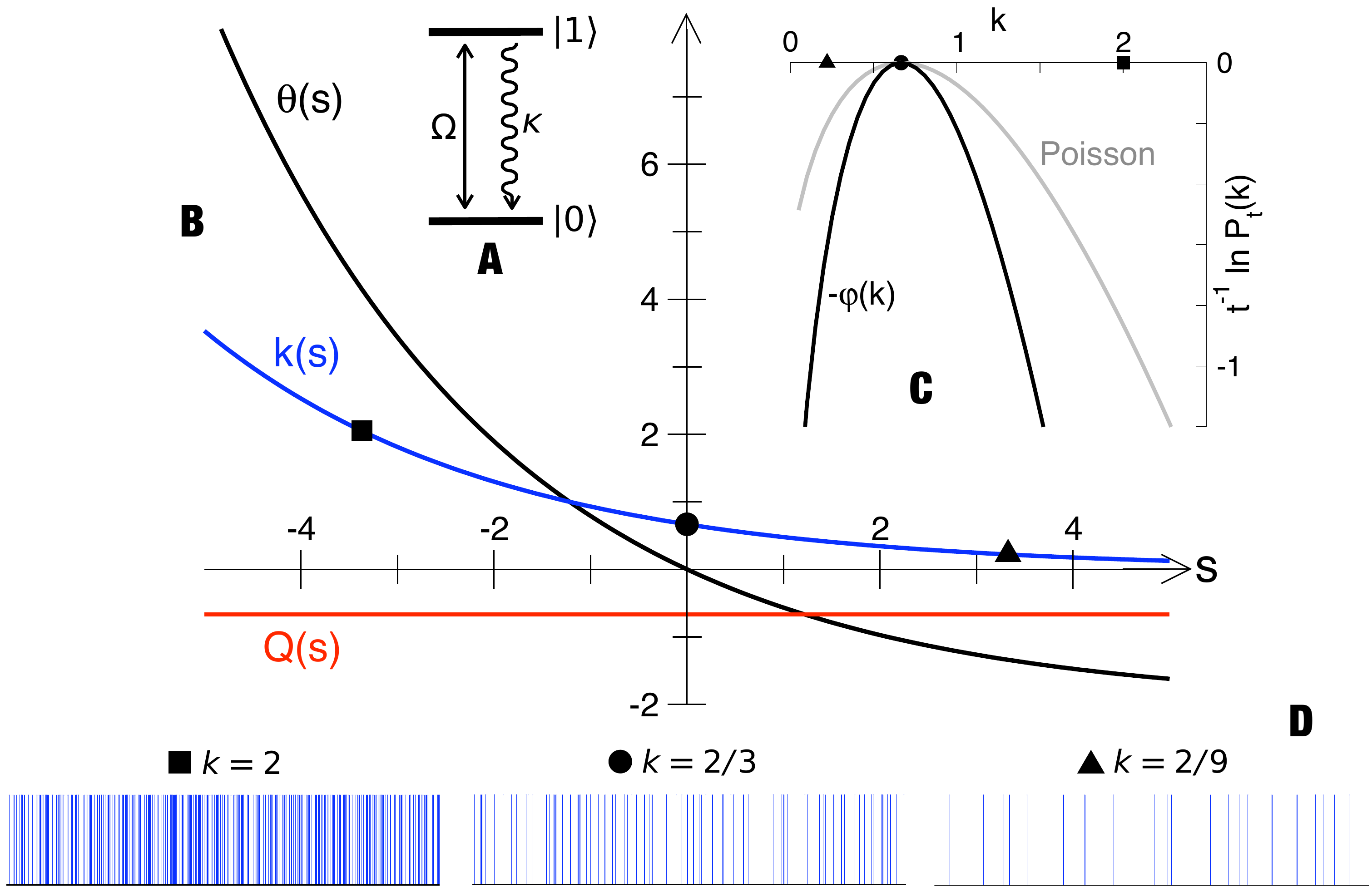}
\caption{
(A) Laser driven 2-level system coupled to a $T=0$ bath.  (B) Large-deviation function $\theta(s)$ of number of emitted photons $K$.  Dynamical trajectories go from more active to less active as $s$, the conjugate field to $K$, is increased, as shown by the average photon rate $k(s) \equiv \langle K \rangle_s/t = - \theta'(s)$.  The Mandel parameter $Q(s) = -2/3$ for all $s$, indicating that for $\kappa = 4 \Omega$ trajectories display a form of scale invariance.   (C) The photon count probability is obtained from (\ref{2level}) by a Legendre transform: $P_t(K) \approx e^{-t \varphi(K/t)}$ with $\varphi(k) = 3 [ k \ln(k/k_0) - (k-k_0)]$.  It is a $\nu=3$ Conway-Maxwell-Poisson distribution \cite{COM}, $P_t(K) \propto [{\rm Poisson}(K;t)]^3$.  (D) Representative trajectories from sub-ensembles with different average $k$.
}
\label{fig1}
\end{figure}

\noindent {\em (i) 2-level system.} Consider a 2-level system, Fig.\ \ref{fig1}A, driven by a resonant laser in contact with a zero temperature bath \cite{Gardiner}.  When the observable $K$ is the number of emitted photons the generalized master operator is
\begin{equation}
{\cal W}_s(\rho) = -i \Omega [ a + a^\dagger , \rho] + e^{-s} \kappa ~ a \rho a^\dagger - \frac{\kappa}{2} \left( a^\dagger a \rho + \rho a^\dagger a \right) ,
\nonumber
\end{equation}
where $a$ and $a^\dagger$ are the lowering and raising operators, $| 0 \rangle \langle 1 |$ and $| 1 \rangle \langle 0 |$, respectively,  $\Omega$ is the Rabi frequency, and $\kappa$ is the decay rate.  We consider the specific choice $\kappa = 4 \Omega$, which is interesting for reasons we discuss below.  Here the LD function takes the simple form
\begin{equation}
\theta(s) = - 2 \Omega \left( 1 - e^{-s/3} \right) ,
\label{2level}
\end{equation}
which is shown in Fig.\ \ref{fig1}B.  It vanishes at $s=0$.  This is a statement of conservation of probability: ${\cal W}_0$ reduces to the master operator of Eq.\ (\ref{Lindblad}) which leaves ${\rm Tr}[\rho]$ invariant.   Derivatives of $\theta(s)$ give moments of the photon number distribution.  In particular, the average number of emitted photons is $k_0 \equiv \langle K \rangle/t = - \theta'(0)$, and the Mandel parameter, $Q_0 \equiv (\langle K^2 \rangle - \langle K \rangle^2)/\langle K \rangle - 1 = -\theta''(0)/\theta'(0)$.  The LD function around $s=0$ encodes the information about fluctuations of {\em typical} trajectories \cite{Lecomte,Brown}.

Away from $s=0$, $\theta(s)$ encodes information about {\em rare} trajectories. Consider the $s$-dependent average photon number (per unit time),
\begin{equation}
k(s) \equiv \frac{\langle K \rangle_s}{t} = \frac{1}{t Z_t(s)} \sum_K K P_t(K) e^{-s K} = -\theta'(s) .
\nonumber
\end{equation}
This expression is the average of $K/t$ where the probability of trajectories is biased by the factor $e^{-s K}$.  Pursuing a thermodynamic analogy, think of $K$ and $s$ as volume and pressure.  Increasing/dereasing pressure leads to a smaller/larger average specific volume, i.e. by controlling pressure we obtain a denser or less dense system.  Something analogous occurs here in the dynamics: $s>0$ corresponds to trajectories with $k(s) < k_0$, i.e. {\em less active} than typical, while $s<0$ corresponds to trajectories with $k(s) > k_0$, i.e. {\em more active} than typical, Fig.\ \ref{fig1}B.

We can also define an $s$-dependent Mandel parameter, $Q(s) \equiv (\langle K^2 \rangle_s - \langle K \rangle_s^2)/\langle K \rangle_s - 1 =-\theta''(s)/\theta'(s) - 1$, which measures the bunching or anti-bunching properties of trajectories with a fixed average photon number $t k(s)$.  For the specific case of $\kappa = 4 \Omega$ we have $k(s) = 2 \Omega e^{-s/3}/3$ (i.e. trajectories go from more to less active as $s$ is increased from negative to positive), but $Q(s) = -2/3$ for all $s$, Fig.\ \ref{fig1}B.   This result is surprising.  We expect photon emissions to be anti-bunched \cite{Gardiner}, but an $s$-independent $Q$ indicates that all sub-ensembles of trajectories, no matter how active or inactive, have the same fluctuation properties of typical trajectories: trajectories would look the same if rescaled by their average emission rate.  Hence $\kappa=4 \Omega$ is a ``special point'' in parameter space where the dynamics displays trajectory scale invariance.  Note that this occurs while all correlation times remain finite.

\begin{figure}
\includegraphics[width=0.8 \columnwidth]{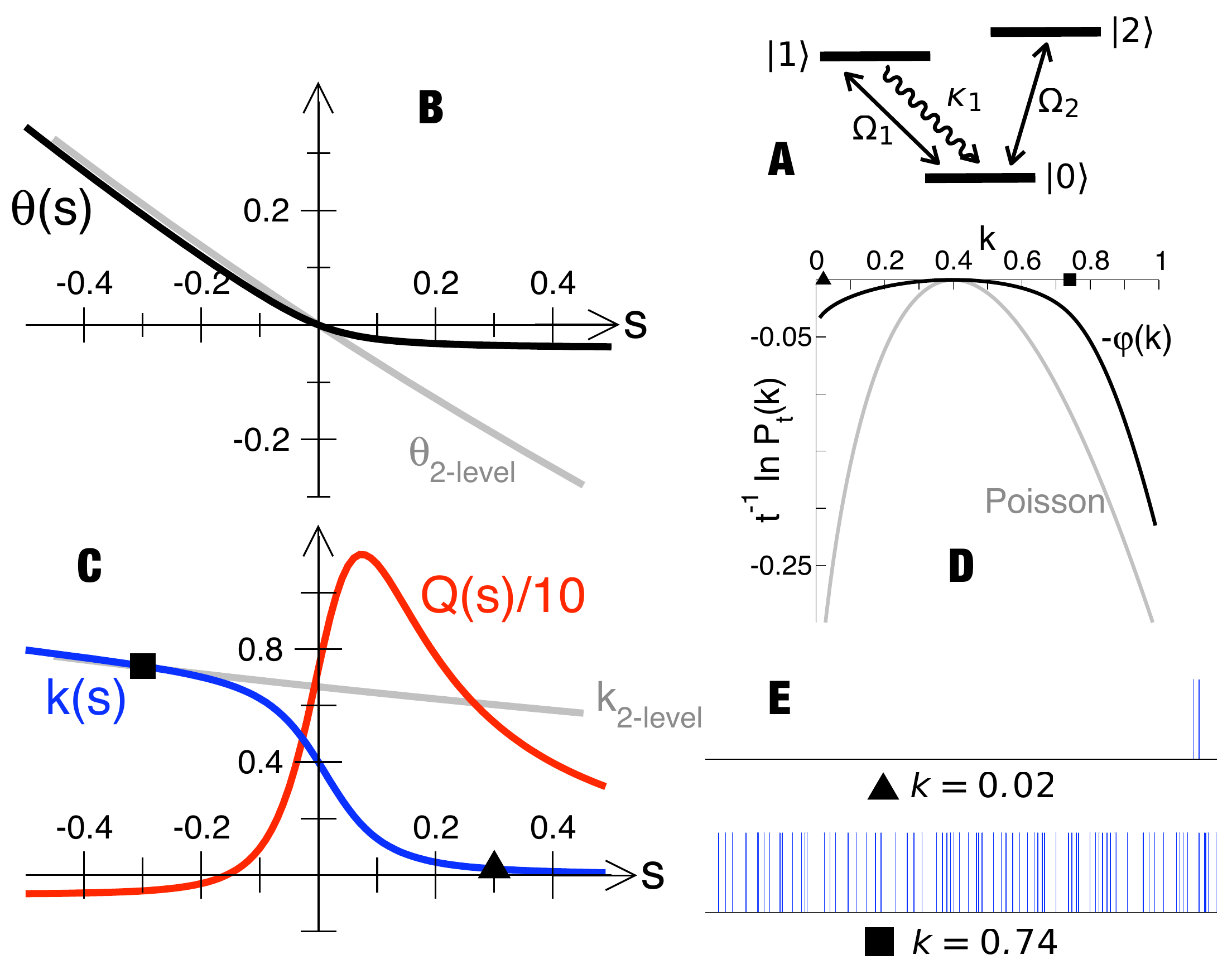}
\caption{(A) Laser driven 3-level system.  Here $\kappa_1 = 4 \Omega_1$ and $\Omega_2 = \Omega_1/10$.   (B,C) The LD function $\theta(s)$ and dynamical order parameter $k(s)$ display crossover behavior near $s=0$ between active and inactive dynamical regimes.  The active side is anti-bunched, $Q<0$.  The inactive side is non-fluctuating $Q=0$.  The peak in $Q$ near $s=0$ signals the dynamical crossover.  (D) The fat tail for $k<k_0$ in $P_t(K)$ is a manifestation of the inactive regime; the thin tail for $k>k_0$ is a manifestation of the active regime.  (E) Representative trajectories from inactive and active sub-ensembles.  At $s=0$ there is (mesoscopic) coexistence of the two dynamical regimes and typical trajectories are intermittent or ``blinking''.
}
\label{fig2}
\end{figure}

\noindent {\em (ii) 3-level system.} Consider now a 3-level system like the one of Fig.\ \ref{fig2}A, driven by two resonant lasers on the $|0\rangle$-$|1\rangle$ and $|0\rangle$-$|2\rangle$ lines with Rabi frequencies $\Omega_1$ and $\Omega_2$, respectively.  Level $|1\rangle$ decays to $|0\rangle$ with rate $\kappa_1$.   We are interested in the statistics of the number $K$ of photons emitted.  When $\Omega_1 \gg \Omega_2$ typical photon emission trajectories are intermittent, displaying ``bright'' and ``dark'' periods \cite{Plenio,Barkai}.  In this case quantum jumps can become evident on macroscopic timescales \cite{Dehmelt}.

The generalized master operator ${\cal W}_s$ is of the form (\ref{Ws}), with $H = \sum_{j=1}^2 \Omega_j (a_j + a_j^\dagger)$, where $a_j \equiv | 0 \rangle \langle j |$ and $a_j^{\dagger} \equiv | j \rangle \langle 0 |$, and only one set of Lindblad terms, $N_{\rm L}=1$, with $L_1 = \sqrt{\kappa_1} a_1$. The LD function $\theta(s)$ is obtained from ${\cal W}_s$ by direct diagonalization.  It is shown in Fig.\ \ref{fig2}B for $\kappa_1 = 4 \Omega_1$ and $\Omega_2 = \Omega_1/10$.   The difference with the 2-level case is striking.  For $s<0$ (i.e. trajectories more active than typical) $\theta(s)$ follow the LD function of the 2-level problem.  Close to $s=0$, however, $\theta(s)$ leaves the 2-level curve and approaches a constant, $\theta(s \gg 0) \approx - \kappa_1 \Omega_2^2$.

Within our thermodynamic analogy this indicates a rapid crossover between two distinct {\em dynamical phases} as we cross $s=0$. Figure \ref{fig2}C shows the corresponding change in $k(s)$.  The active side is that of $s<0$, and trajectories have large $K$.  The inactive side is $s>0$, and trajectories have small $K$.   The active phase is that of the 2-level system $|0\rangle,|1\rangle$ where photon emission is plentiful.  
In the inactive phase 
the atom predominantly occupies the 
$|2\rangle$ state and photon emission is scarce.  The crossover in $k(s)$ is reminiscent of a (smoothed) dynamical first-order transition, such as that seen in the trajectories of certain glassy systems \cite{Garrahan}. The dynamical crossover is also apparent in Mandel parameter, Fig.\ \ref{fig2}C. The active phase is antibunched, $Q(s \ll 0) = -2/3$, while the inactive phase does not fluctuate, $Q(s \gg 0) =0$.  The peak in $Q(s)$ around $s=0$ is a signature of the crossover between phases: here fluctuations are maximal as trajectories are (mesoscopic, i.e. finite time) mixtures of the two coexisting phases. Typical trajectories correspond to $s=0$, but the crossover structure of the LD function $\theta(s)$ has an effect on the tails of the distribution $P_t(K)$, as shown in Fig.\ \ref{fig2}D.  It has a fat tail for $k < \langle K \rangle/t$ [originating from $\theta \lesssim 0$ for $s \gg 0$], and a thin tail for large $k$  [originating from $\theta \approx \theta_{\rm 2-level}$ for $s \ll 0$].

\begin{figure}
\includegraphics[width=0.8 \columnwidth]{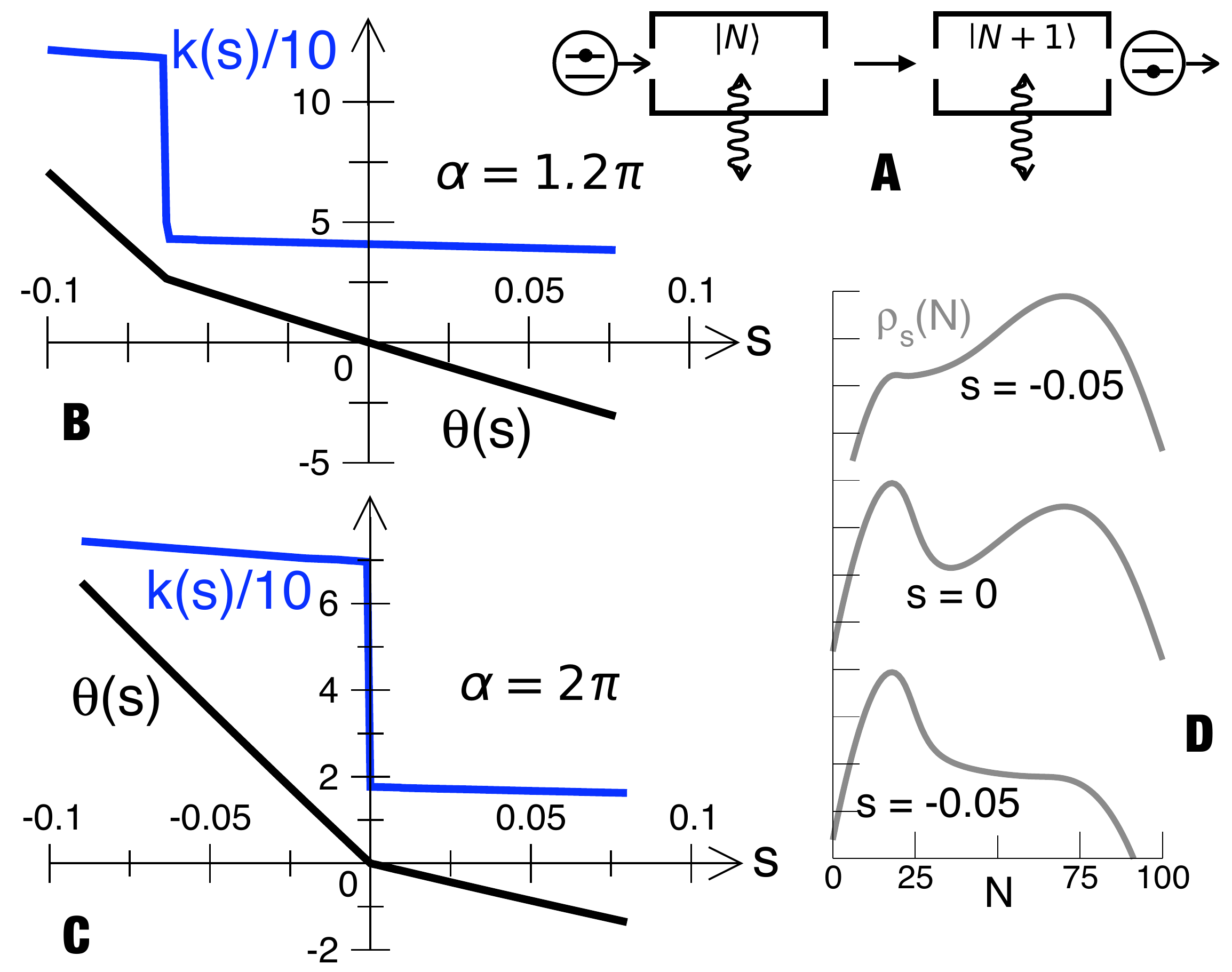}
\caption{Dynamical phase transition in the micromaser.  (A) Cavity mode driven by pumped atoms and interacting with thermal bath.  (B,C) LD function $\theta(s)$ for the number of atomic transitions, $K$.  When the cavity is close to static bistability, $\alpha = 2 \pi$, the LD function has a first-order singularity at $s=0$.  There are two distinct dynamical phases, a more active one with large $K$, and a less active one with small $K$.
Typical trajectories are at coexistence between these phases.   The dynamical transition is still present far from static bistability, $\alpha=1.2 \pi$, but the transition point is at $s<0$, i.e. dynamical coexistence will be only manifest in rare trajectories.   (D) Cavity photon distribution in active and inactive phases, and at coexistence (i.e. stationary density matrix). }
\label{fig3}
\end{figure}

\noindent {\em (iii) Micromaser.} We now consider the problem of a micromaser \cite{Englert}, a resonant cavity coupled to a finite temperature bath and pumped by excited two level atoms which are sent into the cavity with a constant rate, Fig.\ \ref{fig3}A.  The atoms only interact with a single mode of the cavity.  The cavity reaches a steady state which is sensitive on pump rate and atom-cavity coupling.  In particular, the steady state of the cavity can change from unimodal to bimodal \cite{Englert}.  We now show that this static bistability has an associated dynamic bistability.

Our dynamical order parameter $K$ is now the number of atoms which leave the cavity and are in the ground state.  The super-operator ${\cal W}_s$ (\ref{Ws}) follows from the Lindblad master equation for the cavity after tracing out the atom and the thermal bath \cite{Englert}.  There are four sets of Lindblad operators, $N_{\rm L}=4$, two from the atom-cavity interaction, $L_1 = \sqrt{r} \frac{\sin \left( \phi \sqrt{aa^\dagger} \right)}{\sqrt{aa^\dagger}} a$ and $L_2 = \sqrt{r} \cos \left( \phi \sqrt{aa^\dagger} \right)$, and two from the cavity-bath interaction, $L_3 = \sqrt{\kappa} a$ and $L_4 = \sqrt{\lambda} a^\dagger$.  Here $a,a^\dagger$ are the raising/lowering operators of the cavity mode, $r$ is the atom beam rate, $\kappa$ and $\lambda$ are the thermal relaxation and excitation rates, and $\phi$ encodes the atom-cavity interaction \cite{Englert}. Events are recorded when quantum jumps under the action of $L_1$ occur.

The LD function $\theta(s)$ can be obtained by assuming that the corresponding eigenmatrix $r_s$ of ${\cal W}_s$ (see below) is diagonal in $a^\dagger a$.   It is shown in Figs.\ \ref{fig3}B,C for two values of the ``pump parameter'' $\alpha \equiv \phi \sqrt{r/(\kappa-\lambda)}$ \cite{Englert}.
For $\alpha=2\pi$ the stationary state of the cavity is close to being bistable, undergoing a sudden change from a low average photon occupation $\langle N \rangle$ at $\alpha \lesssim 2 \pi$ to a large $\langle N \rangle$ at $\alpha \gtrsim 2 \pi$ \cite{Englert}. In this case the LD function is singular at $s=0$, and $k(s)$ has a discontinuous jump, Fig.\ \ref{fig3}C. This is a dynamic, or space-time, phase transition \cite{Garrahan}.  It is first-order because the order parameter, $k$, changes discontinuously.  The phase transition is between an active phase at $s<0$ and an inactive (or less active) one at $s>0$.  Fig.\ \ref{fig3}D shows that the active phase corresponds to that of large $\langle N \rangle$ and the inactive one to small $\langle N \rangle$.  The transition point is $s=0$ so that normal dynamics occurs under {\em dynamic phase-coexistence}.  The dynamical transition remains even far from static bistability, but the transition point moves away from $s=0$, Fig.\ \ref{fig3}B.

\noindent {\em Mapping of rare trajectories to typical ones.} The LD function $\theta(s)$ encodes properties of rare quantum trajectories, and by exploiting the analogy with thermodynamics we can describe sub-ensembles of trajectories as dynamical or space-time phases \cite{Garrahan}.  $\partial_t \rho = {\cal W}_s(\rho)$ however is not a physical time evolution, but we can show \cite{future} that there is an alternative trace-preserving evolution which generates the same $s$-ensemble.  Thus, rare trajectories in one system correspond to typical trajectories of a related system, and crossovers or transitions controlled by $s$, such as the ones discussed above, can be realized as transitions controlled by physical parameters. 

The super-operator ${\cal W}_s$ has the LD function $\theta(s)$ as its largest real eigenvalue, with ``right'' and ``left'' Hermitian eigenmatrices $r_s$ and $l_s$, respectively \cite{left}. These eigenmatrices obey ${\cal W}(r_s) = \theta(s) r_s$ and $(l_s){\cal W}_s = l_s \theta(s)$ and we normalize them such that ${\rm Tr}[r_s]=1$ and ${\rm Tr}[l_s]={\rm Tr}[{\mathbb I}]$.  Given a matrix $\rho_s(t)$ which evolves according to $\partial_t \rho_s = {\cal W}_s(\rho_s)$, there is an associated density matrix $\tilde{\rho}(t) \equiv  l_s^{1/2} \rho_s(t) l_s^{1/2} / {\rm Tr}[l_s \rho_s(t)]$, whose corresponding evolution, $\partial_t \tilde{\rho} = \tilde{\cal W}(\tilde{\rho})$, is of the Lindblad form (\ref{Lindblad}), with the following Hamiltonian and Lindblad operators:
\begin{eqnarray}
\tilde{H} &=& \frac{1}{2} l_s^{-1/2} \left( \{ H , l_s \} +\frac{i}{2} [ L_\mu^\dagger L_\mu , l_s ] \right) l_s^{-1/2} ,
\label{Htilde}
\\
\tilde{L}_\mu &=& \left[ \delta_{\mu 1} e^{-s/2} + (1- \delta_{\mu 1}) \right] l_s^{1/2} L_\mu l_s^{-1/2} .
\label{Ltilde}
\end{eqnarray}
This dynamics is trace-preserving, $({\mathbb I})\tilde{{\cal W}} = 0$, and the set of trajectories of quantum jumps due to $\tilde{L}_1$ coincides with the $s$-ensemble of ${\cal W}_s$.  The tilde process is that of a physical dynamics.  Its typical trajectories correspond to rare trajectories of the original process ${\cal W}$ \cite{Evans}.

The explicit construction of the trace-conserving system gives interesting insights into the structure of trajectories away from $s=0$ \cite{future}: (i) For the 2-level system above we have $\tilde{H} = e^{-s/3} \Omega  (\tilde{a} + \tilde{a}^\dagger)$ and $\tilde{L}_1 =  e^{-s/6} \sqrt{\kappa} \tilde{a}$.  This is of the same form as the $s=0$ problem with all rates multiplied by $e^{-s/3}$.  That is, rare trajectories ($s \neq 0$) are typical trajectories of the same system but with time rescaled as $t \to e^{s/3} t$.  This confirms $\kappa = 4 \Omega$ as a special symmetry point of the 2-level problem.  (ii) For the 3-level system above on the inactive side, $s > 0$, the mapped system is another 3-level problem with an additional strong laser on the $| \tilde{1} \rangle$-$| \tilde{2} \rangle$ line.  This coupling introduces an effective detuning for the laser on the $| \tilde{0} \rangle$-$| \tilde{1} \rangle$ transition, suppressing the excitation of state $| \tilde{1} \rangle$ and subsequent photon emission \cite{future}.

The ``statistical mechanics of trajectories'' method presented here seems to be a useful framework to study non-equilibrium dynamics in open quantum systems.  Already for very simple problems it reveals unanticipated richness.  We expect it to be even more fruitful in the study of non-equilibrium quantum many-body systems.

We thank Andrew Armour and Markus M\"uller for helpful discussions.

\end{document}